\documentclass[conference]{IEEEtran}
\IEEEoverridecommandlockouts
\usepackage{cite}
\usepackage{amsmath,amssymb,amsfonts}
\usepackage{algorithmic}
\usepackage{graphicx}
\usepackage{textcomp}
\usepackage{soul}
\usepackage{xcolor}
\def\BibTeX{{\rm B\kern-.05em{\sc i\kern-.025em b}\kern-.08em
    T\kern-.1667em\lower.7ex\hbox{E}\kern-.125emX}}
\usepackage{fancyhdr}
\begin{document}

\title{A 0.74-mW K$_{\mathrm{u}}$-band Cryogenic SiGe LNA in 130-nm BiCMOS with 14-K Noise Temperature for Scalable Quantum Readout 
\thanks{*: equal contribution}
}

\author{
\IEEEauthorblockN{
Xiuhan Chen\textsuperscript{*1}, 
Haoling Li\textsuperscript{*1}, 
Gun Suer\textsuperscript{1}, 
Leonardo Ranzani\textsuperscript{2},
Kin Chung Fong\textsuperscript{1},
Aravind Nagulu\textsuperscript{1},
Najme Ebrahimi\textsuperscript{1}
}

\vspace{1em}

\IEEEauthorblockA{\textsuperscript{1}\textit{Northeastern University, Boston, MA, USA }}

\IEEEauthorblockA{\textsuperscript{2}\textit{RTX BBN Technologies, Cambridge, MA, USA }}
}

\maketitle
\thispagestyle{fancy}

\begin{abstract}
This paper presents a sub-milliwatt K$_{\mathrm{u}}$-band cryogenic low-noise amplifier (LNA) implemented in GlobalFoundries (GF) 130CBIC SiGe BiCMOS, intended for a sub-Kelvin qubit readout chain integrating a JPA and SiGe LNA on the Still stage. The four-stage cascode LNA features balun-coupled interstage matching with a Q-enhanced resonance tank controlled by 2-bit switches tank for frequency tuning and a voltage-controlled cross-coupled negative-$g_m$ cell. The operating point of the Q-cell is optimized at each temperature stage to achieve the best in-band performance under a stable condition. Characterized at 2.5\,K for the initial demonstration, the LNA achieves an average noise temperature of $\sim$14\,K across 12--18\,GHz and 28\,dB peak gain around 16.5\,GHz, while consuming only \textbf{0.74\,mW}. To the authors' knowledge, this is the lowest reported DC power for a K$_{\mathrm{u}}$-band cryogenic SiGe LNA.
\end{abstract}


\section{Introduction}
\begin{figure}[t]
\centering
\includegraphics[width=0.65\columnwidth]{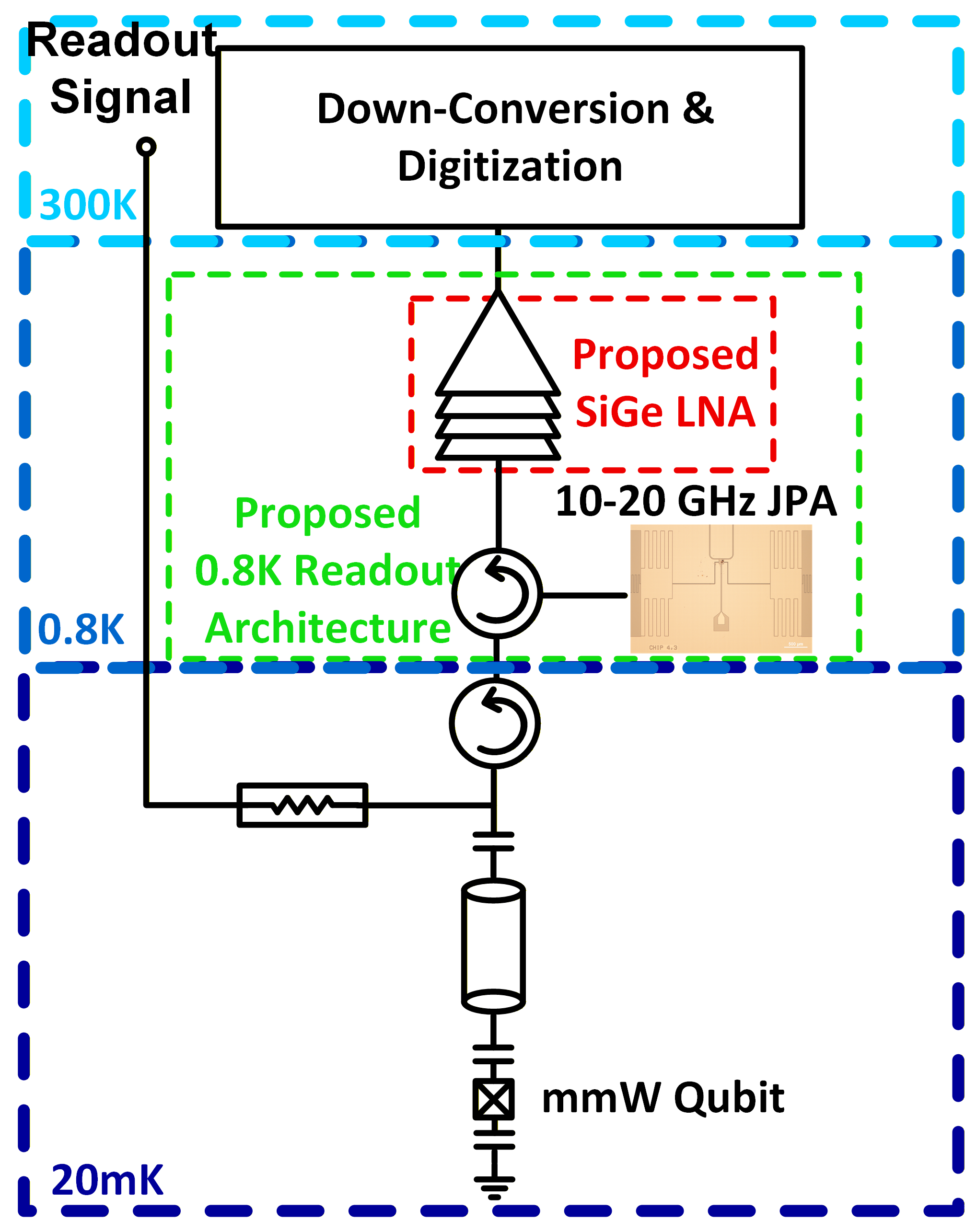}
\caption{Proposed Still-stage ($\sim$0.8\,K) quantum readout architecture. A chain of two circulators isolates the source from the reflective Josephson parametric amplifier (JPA) and routes the amplified signal to the proposed SiGe LNA. Both the JPA and the LNA are integrated and placed on the Still stage of the dilution refrigerator. The amplified signal is sent off to the room-temperature down-conversion and digitization chain, abstracted as a single block.}
\label{fig:arch}
\end{figure}

Scaling superconducting quantum processors operating at 10-100\,mK is expected to scale to thousands of qubits. However, the scaling is bottlenecked by (i) the dense coaxial cables connecting the different cryogenic stages inside a dilution refrigerator, and (ii) the rapid reduction in available cooling power with stage temperature, approximately following a \(T^{-4}\) dependence\cite{krinner_engineering_2019}\cite{CMOS_Quantum_System}\cite{arute_quantum_2019}\cite{Review_quangtum}. To overcome these limitations, a promising approach is proposed that moves the first-stage JPA\cite{JPA}\cite{JPA_classical} from the millikelvin (mK) stage to the Still stage of the dilution refrigerator ($\sim$0.8\, K) and integrates it with the second-stage SiGe LNA on the same printed circuit board (PCB).  The integration placement requires the JPA remain superconducting in the 1--4\, K range, which excludes conventional Al-based JPAs ($T_c\!\approx\!1.2$\, K) but is satisfied by proposed van der Waals NbSe$_2$/WSe$_2$/NbSe$_2$ Josephson junctions ($T_c\!\approx\!7$\,K)\cite{NbSe2}. Besides, the readout frequency should ideally satisfy $\hbar\omega \gg k_B T$  at the Still-stage operating temperature to suppress thermal-photon occupancy, leading to a frequency shift from the conventional 4--8\,GHz band into the K$_{\mathrm{u}}$-band ($\sim$12--18\,GHz) or above. Consequently, realizing a K$_{\mathrm{u}}$-band sub-Kelvin integrated readout system as shown in Fig.~\ref{fig:arch} can be a potential scaling solution. The current work develops the cryogenic K$_{\mathrm{u}}$-band SiGe LNA and the integration of SiGe LNA with JPA will be demonstrated as a part of future work.

\begin{figure*}[t]
    \centerline{\includegraphics[width=0.9\textwidth]{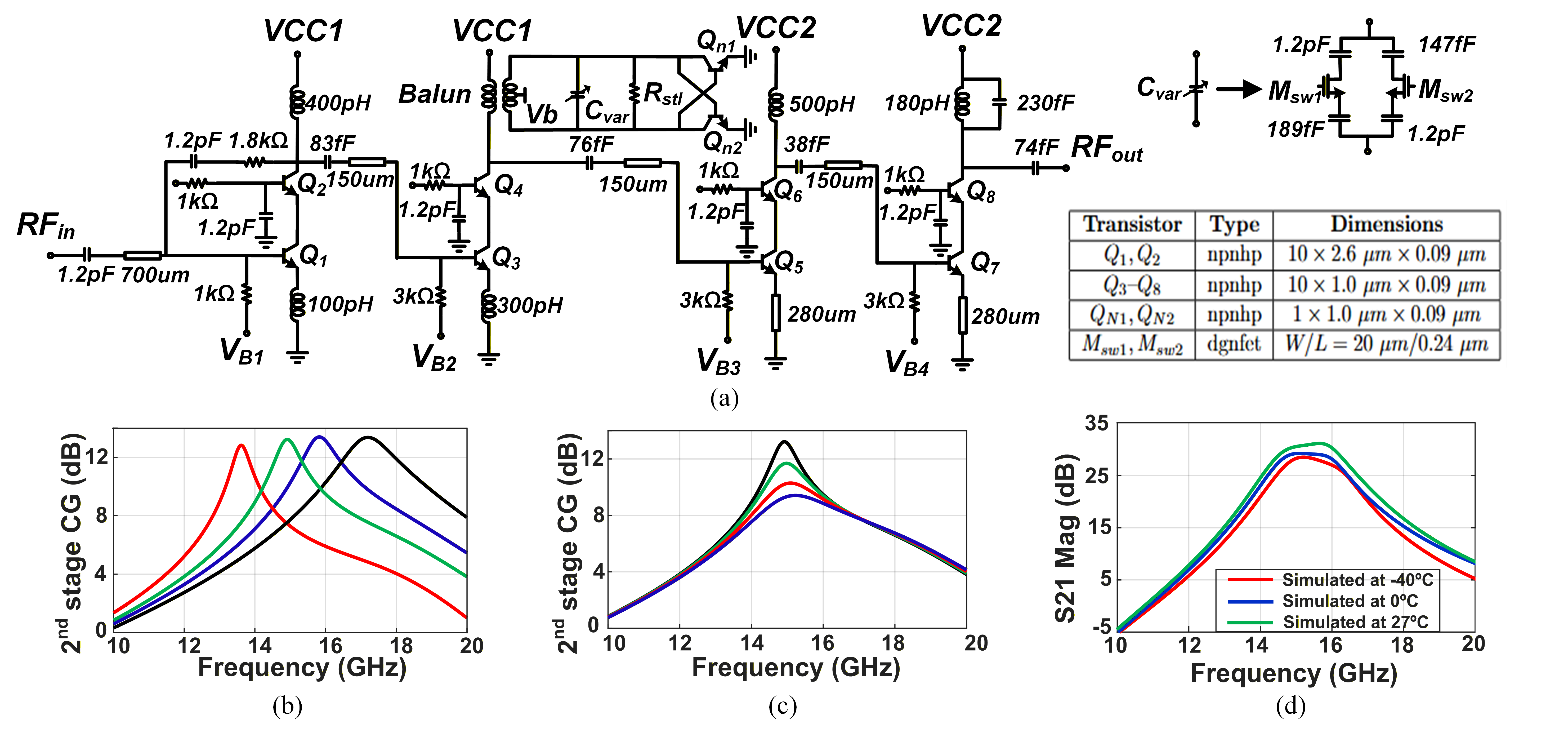}}
    \caption{(a)~Schematic of the four-stage cascode amplifier with components' values annotated. The right inset shows the implementation of the 2-bit varactor $C_{\mathrm{var}}$ using two switched MOSFET capacitor branches $M_{sw1}/M_{sw2}$ and HBT/MOSFET device dimensions. (b)~Simulated second-stage $S_{21}$ at 300\,K for the four $C_{\mathrm{var}}$ codes (with $-g_m$ bias fixed), showing coarse frequency tuning across $\sim$13.5--17\,GHz with peak gain of $\sim$13\,dB. (c)~Simulated second-stage $S_{21}$ at 300\,K under sweep of the $-g_m$ bias $V_b$ (with $C_{\mathrm{var}}$ fixed), showing fine gain trim from $\sim$9\,dB to $\sim$13\,dB. (d)~Simulated $S_{21}$ of the full LNA at the temperatures $-40^\circ\mathrm{C}$, $0^\circ\mathrm{C}$, and $27^\circ\mathrm{C}$ where the PDK model is available.}
    \label{fig:schematic}
\end{figure*}

The required LNA must operate at K$_{\mathrm{u}}$ band with ultra-low DC power, so that scaling to thousands of readout channels satisfies the Still cooling budget\cite{krinner_engineering_2019}. SiGe HBTs are attractive for this application, because the proportional cryogenic increase in $g_m$ and $f_T$ benefits noise performance, and the BiCMOS platform enables CMOS co-integration. Sub-mW SiGe cryogenic LNAs have been demonstrated with $T_n$ of 2--5\,K below 8\,GHz\cite{ref_zou2024}, however, demonstrations of sub-mW K$_{\mathrm{u}}$-band SiGe cryogenic LNAs remain exceedingly rare.

This work targets that gap in GlobalFoundries 130CBIC, a recently released SiGe BiCMOS platform with NPN $f_T/f_{\max}$ exceeding 400\,GHz. Due to the lack of cryogenic model of this newly released technology, the potentially shifted resonances and instability of any gain-peaking circuitry need to be carefully considered. The proposed LNA addresses this through including a second-stage Q-enhanced tank with 2-bit switched-capacitor tuning and an independently biased $-g_m$ cell for re-optimization at each temperature. The remainder of this paper will focus on the analysis of the schematic and measurement results.

\section{Circuit implementation and topology}

Fig.~\ref{fig:schematic}(a) depicts the schematic of the proposed four-stage cascode LNA. All stages are supplied by 0.7\,V DC sources and biased by a common current-mirror reference, with stages~1--2 sharing the VCC1 rail and stages~3--4 sharing VCC2. The input stage uses a cascode pair (Q$_1$/Q$_2$,$10\!\times\!2.6\,\mu\mathrm{m}\!\times\!0.09\,\mu\mathrm{m}$) biased at its minimum-NF current density and a 100\,pH emitter degeneration inductor at the emitter of Q$_1$ to enable simultaneous noise and impedance matching. An on-chip transmission line and a series capacitor are utilized here to match the noise and impedance to 50$\Omega$. To widen the input bandwidth and increase the stability, a shunt feedback network ($R_f$~=~1.8\,k$\Omega$, $C_f$~=~1.2\,pF) is placed from the Q$_2$ collector back to the Q$_1$ base. The feedback also reduces the sensitivity of the input match to device-parameter drift due to the temperature changes.

Since no 4\,K model is available for GF 130CBIC, design margins were established from (i) PDK-temperature $S_{21}$ simulations shown in Fig.~\ref{fig:schematic}(d), and (ii) prior cryogenic characterizations of 130-nm SiGe BiCMOS \cite{ref_SiGe_LNA_review, cryo_bjt}, reporting 2--3$\times$ $g_m$ enhancement and $Q$ improvement. 

The second-stage (Fig.~\ref{fig:schematic}(a)) utilizes the technique of a tunable, Q-enhanced resonance tank that enables frequency tunability and performance robustness at cryogenic temperatures. It combines an on-chip integrated balun that primarily serves as the collector inductor of $Q_3$/$Q_4$, a fixed parallel damping resistor $R_{\mathrm{stl}}$ at the balun secondary, 2-bit varactor $C_{var}$ implemented as switched MOSFET capacitor branches, and a cross-coupled negative-$g_m$ pair $Q_{n1}/Q_{n2}$, independently biased through $V_b$. The increase in the quality factor of passive components at cryogenic temperatures will cause the quality factor of the resonance tank to be higher than its room temperature design value, and will compress the stability margin of any gain peak circuit. Therefore, the resistor $R_{stl}$ is introduced to intentionally dampen the resonant tank circuit, ensuring it to fall back to the predetermined load Q value and the stability budget is not affected by temperature changes. However, relying on $R_{stl}$ alone will result in a loss of peak gain under all temperature conditions. To address this, a pair of cross-coupled transistors introduces a negative conductance of $-g_{m,n}/2$ to the tank which increases the $Q$ of the resonant tank circuit, and compensates for the lost gain. The effective tank conductance is
\begin{equation}
G_{\mathrm{eff}} \;=\; G_{\mathrm{tank}} \;+\; 1/R_{\mathrm{stl}}\;-\; g_{m,n}/2 \;>\; 0,
\label{eq:gtot}
\end{equation}
where the strict positivity of $G_{\mathrm{eff}}$ is the small-signal stability condition. The corresponding tank voltage gain at resonance is
\begin{equation}
A_{v,\mathrm{tank}} \;=\; \frac{g_{m,3}}{G_{\mathrm{eff}}}
\;=\; \frac{g_{m,3}}{G_{\mathrm{tank}}+1/R_{\mathrm{stl}}-g_{m,n}/2},
\label{eq:av}
\end{equation}

in which $g_{m,3}$ is the transconductance of the common-emitter HBT $Q_3$. Increasing $|g_{m,n}|$ raises the gain but pushes $G_{\mathrm{eff}}$ toward zero. Independent control of $V_b$ allows the $-g_m$ bias to be re-optimized at each operating temperature so that $G_{\mathrm{eff}}$ remains positive while the in-band gain is maximized.

Fig.~\ref{fig:schematic}(b) shows the simulated second stage gain at 300\,K for the four $C_{\mathrm{var}}$ codes with $V_b$ fixed. The resonance peak shifts across $\sim$13.5--17\,GHz with peak gain of $\sim$12\,dB, providing a coarse frequency-tuning mechanism to absorb the cryogenic frequency drift caused by junction-capacitance and transmission-line phase-velocity shifts. Fig.~\ref{fig:schematic}(c) shows the simulated second stage gain at the same operating point with $C_{\mathrm{var}}$ fixed and $V_b$ swept. The peak gain and the bandwidth change at different bias points, confirming the effects of the $-g_m$ cell on the second stage. The simulated $S_{21}$ across the available temperature range of the GF 130CBIC PDK is presented in Fig.~\ref{fig:schematic}(d). While the enhanced $g_m$ and $Q$ provide higher peak gain, the temperature-induced frequency shift of each stage results in a trend of broader bandwidth for the LNA.

\section{Measurement Setup and Results}

\begin{figure}[t]
\centering
\includegraphics[width=\columnwidth]{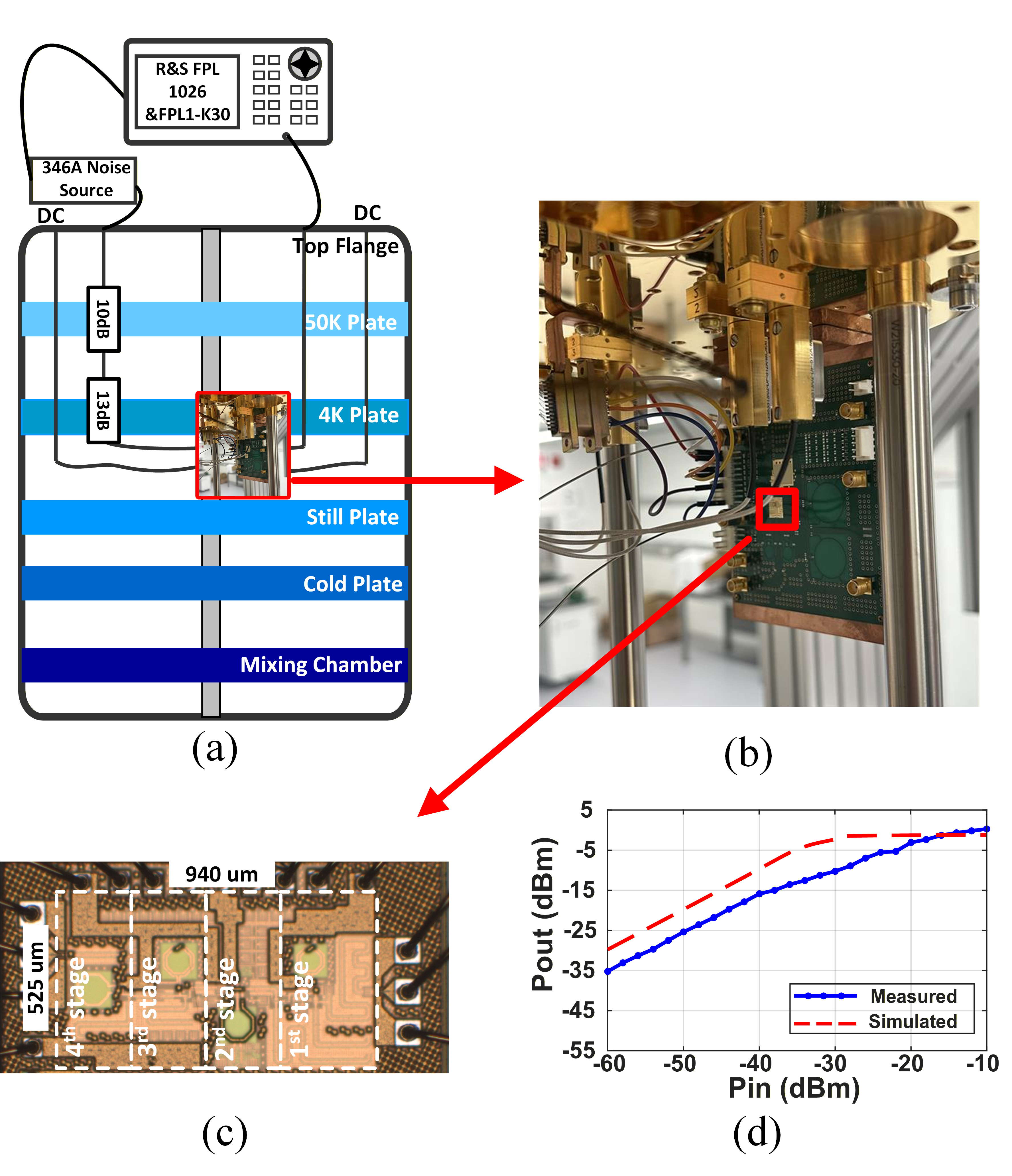}
\caption{(a) Diagram of the cryogenic noise figure measurement setup. (b) Photograph of the chip-on-PCB assembly mounted to the 4 K plate of the Bluefors dilution refrigerator used for cryogenic characterization. (c) Die micrograph of the fabricated chip. The four-stage K$_{\mathrm{u}}$-band cryogenic LNA core is outlined in white. (d) Measured and simulated $P_{\mathrm{out}}$ versus $P_{\mathrm{in}}$ of LNA at 15\,GHz}
\label{fig:setup}
\end{figure}

The 4-stage LNA was fabricated in GlobalFoundries 130\,nm CBIC SiGe BiCMOS, occupying a core area of $0.525\,\mathrm{mm} \times 0.940\,\mathrm{mm}$. The chip micrograph is shown in Fig.~\ref{fig:setup}(c) and the 4-stage LNA is highlighted by the white outline.

For both room-temperature and cryogenic characterization, the die was wire-bonded onto a dedicated RF PCB fabricated on Rogers RO4350 substrate with a non-magnetic ENEPIG surface finish. $S$-parameters were measured with a Rohde \& Schwarz ZNB3020 vector network analyzer, and noise figure was measured with a Rohde \& Schwarz FPL1026 spectrum analyzer embedded with the FPL1-K30 noise figure measurement kit, using the Y-factor method with an AT-346A noise source (10\,MHz--18\,GHz). The upper frequency limit of the noise source restricts noise-figure data to $\leq$18\,GHz.

 Cryogenic measurements were performed in a Bluefors dilution refrigerator with the PCB anchored to the 4\,K plate (Fig.~\ref{fig:setup}(a)–(b)), where the measured physical temperature was 2.5\,K. DC bias used phosphor-bronze looms with in-line RC filtering and RF I/O used thermally-anchored semi-rigid coaxial cables. Cable and PCB-trace losses were de-embedded from the measured data using datasheet specifications and EM simulations. An additional 3–4\,K is included in the reported noise temperature to account for the temperature gradient along the cryogenic coaxial cables.

\begin{figure}[!t]
    \centering
    \includegraphics[width=0.75\columnwidth]{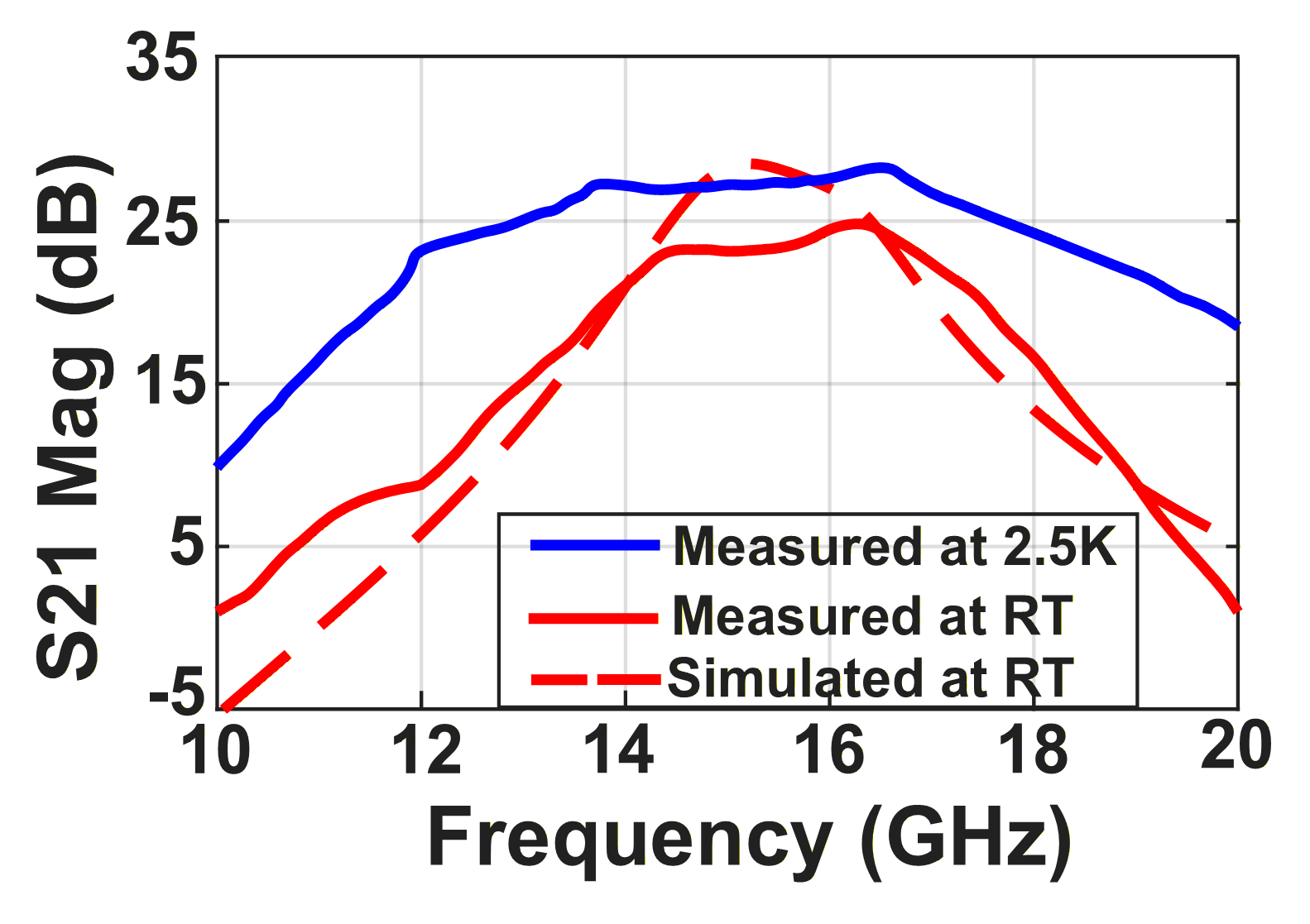}
    \caption{Measured small-signal gain $|S_{21}|$ of the LNA across 10--20\,GHz. The measured peak gain is 25\,dB at $\sim$16\,GHz at 300\,K (red) and rises to 28\,dB near 16.5\,GHz at 2.5\,K (blue). The dashed red curve shows the 300\,K simulation.}
    \label{fig:s21}
\end{figure}

\begin{figure}[!t]
    \centering
    \includegraphics[width=0.75\columnwidth]{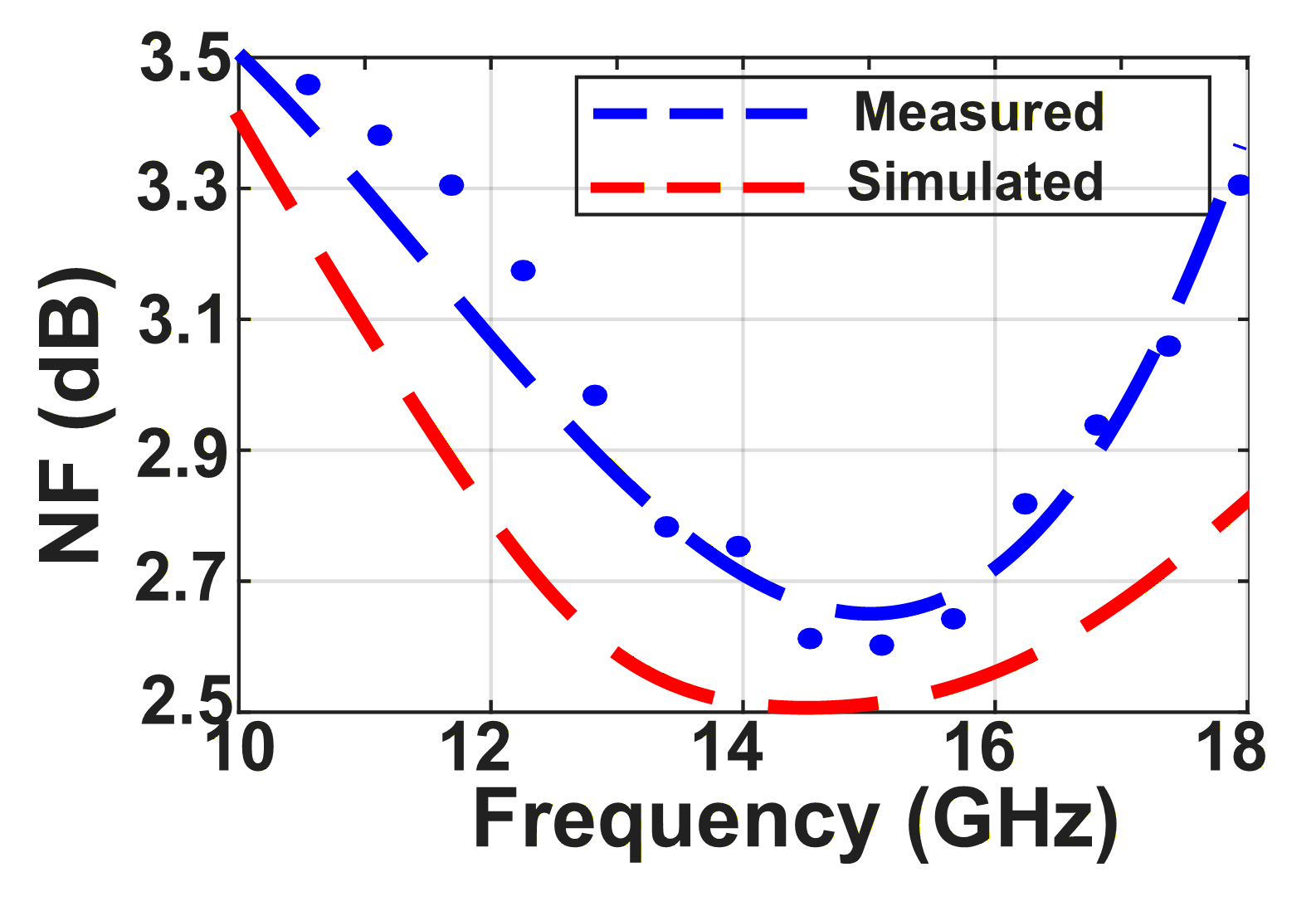}
    \caption{Measured (blue) and simulated (red) noise figure of the LNA at 300\,K across 10--18\,GHz. The minimum measured NF is approximately 2.65\,dB near 15\,GHz, in good agreement with the simulated minimum of $\sim$2.5\,dB.}
    \label{fig:nf_rt}
\end{figure}

Fig.~\ref{fig:s21}~and~\ref{fig:nf_rt} present the room-temperature measured $S_{21}$ magnitude and noise figure, together with the corresponding simulations. The measured $S_{21}$ exhibits a peak gain of 25 dB at 16 GHz, with a total DC power consumption of 5.2 mW at 300 K, which shows a good agreement with simulation. The slight resonance frequency shift due to parasitic causes the measured $S_{21}$ to have a lower gain but wider bandwidth. The measured minimum noise figure is 2.65\,dB at approximately 15\,GHz, which aligns well with the simulated minimum value of 2.5\,dB. Fig.~\ref{fig:setup}(d) shows the measured and simulated $P_{\mathrm{out}}$ versus $P_{\mathrm{in}}$ at 15~GHz, indicating an input-referred $P_{\mathrm{1dB}}$ of approximately $-38$~dBm.

\begin{figure}[!t]
    \centering
    \includegraphics[width=0.75\columnwidth]{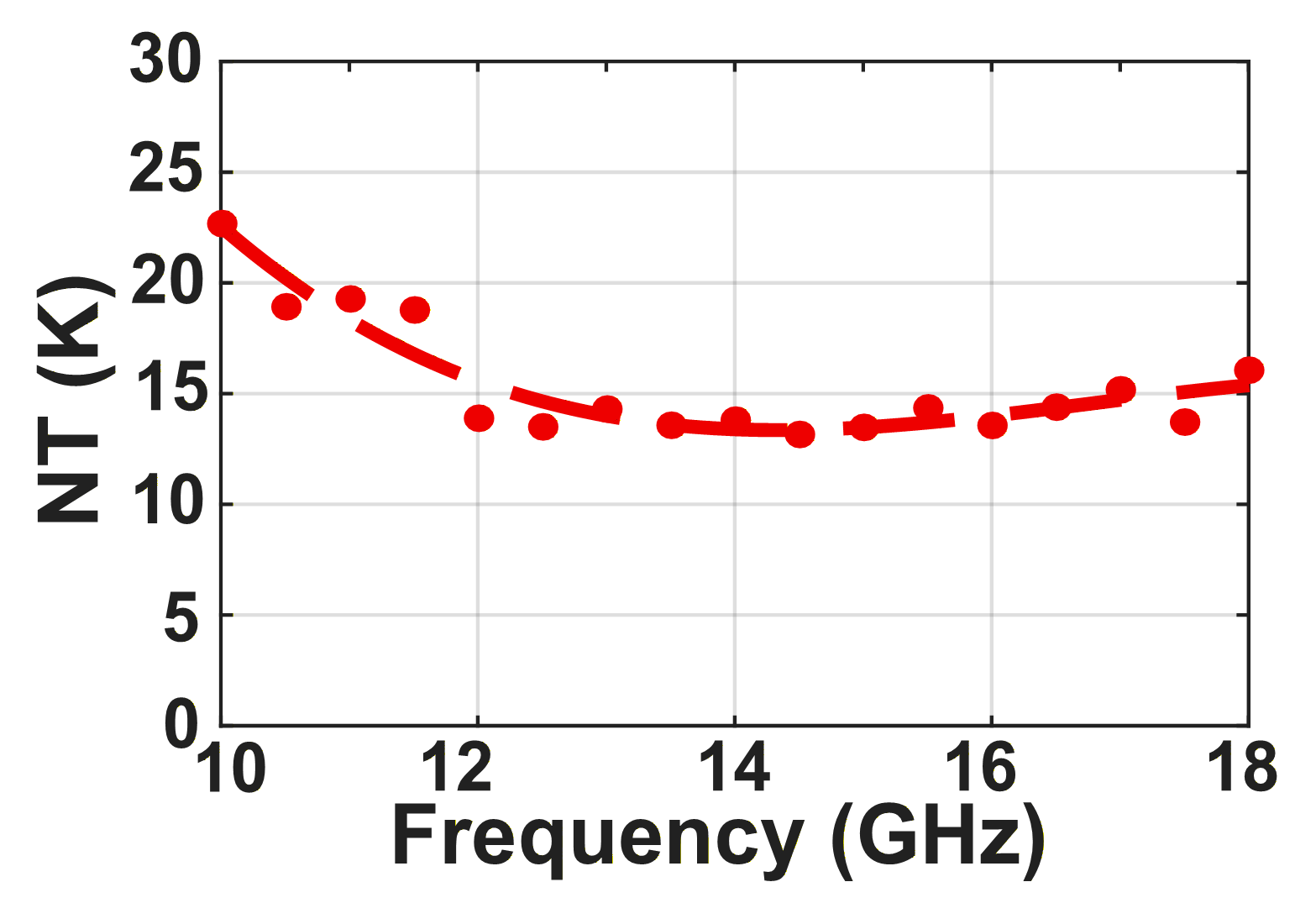}
    \caption{Measured noise temperature $T_n$ of the LNA at 2.5\,K across 10--18\,GHz. $T_n$ achieved an average value 14\,K across the 12--18\,GHz band, with a minimum of $\sim$13\,K near 15\,GHz}
    \label{fig:nt_cryo}
\end{figure}

Fig.~\ref{fig:s21}~and~\ref{fig:nt_cryo} show the measured $S_{21}$ and noise temperature at 2.5\,K. As described in Section~II, the DC bias is re-optimized for this cryogenic operating condition. The total current flowing through the LNA is 1.06\,mA and the DC power consumption is reduced to 0.74\,mW, corresponding to a $7\times$ decrease compared to the room-temperature operating point, primarily due to the enhanced $g_m$ and reduced thermal loss at cryogenic temperature. At 2.5\,K, the measured peak $S_{21}$ rises to 28\,dB and the frequency of the peak gain shifts slightly upward to $\sim$16.5\,GHz. The increase in peak gain and bandwidth is attributed to the cryogenic enhancement of the HBT $g_m$ and the frequency shift of each stage, extending the same trend observed in the PDK temperature sweep of Fig.~\ref{fig:schematic}(d). The second-stage stability resistor $R_{\mathrm{stl}}$ and the re-optimized $-g_m$ bias jointly keep a positive $G_{\mathrm{eff}}$ margin at this operating point. The measured noise temperature exhibits an average of 14\,K across the 12--18\,GHz band. The cryogenic noise temperature represents an approximately $18\times$ reduction relative to the 300\,K minimum NF of 2.65\,dB. Large-signal characterization at 2.5\,K was attempted but is not reported here. Since the increased cryogenic $g_m$ tightens the small-signal stability margin under large-signal excitation, oscillation was observed before reaching the compression point. Adjustments to $R_{\mathrm{stl}}$ and $-g_m$ bias and other techniques for stability improvement to provide larger headroom margin are noted as future work.

\begin{table}[t]
\centering
\caption{Comparison with state-of-the-art cryogenic SiGe LNAs.}
\label{tab:comparison}
\renewcommand{\arraystretch}{1.15}
\setlength{\tabcolsep}{3pt}
\resizebox{\columnwidth}{!}{%
\begin{tabular}{l c c c c c c}
\hline\hline
\textbf{Ref.} & \textbf{Tech.} & \textbf{Freq.} & \textbf{Gain} & \textbf{$T_n$} & \textbf{$P_{\mathrm{DC}}$} & \textbf{$T_{\mathrm{phys}}$} \\
 & \textbf{(nm)} & \textbf{(GHz)} & \textbf{(dB)} & \textbf{(K)} & \textbf{(mW)} & \textbf{(K)} \\
\hline
Wu~\cite{ref_wu2023}          & 130 SiGe & 2--12    & 43$^{*}$       & 6$^{*}$       & 1.3  & 4 \\
Wong~\cite{Wong_2016} & 130 SiGe & 22 &  25  &   35   &  0.9  &   15  \\  
Montazeri~\cite{ref_montazeri2017} & 130 SiGe & 4--8     & 30/26$^{*}$   & 7.6/8$^{*}$       & 0.76/0.58 & 18  \\
Zou~\cite{ref_zou2024}        & 130 SiGe & 3--6     & $>$\,30$^{\dagger}$ & 2.6$^{*}$     & 1.6  & 7   \\
Ramírez~\cite{ref_varonen2019}& 130 SiGe & 52--65   & $>$\,15.6$^{\dagger}$    & 191$^{*}$     & 6.3  & 20  \\
\hline
\textbf{This work}                 & \textbf{130 SiGe} & \textbf{12--18} & \textbf{28$^{\ddagger}$} & \textbf{14$^{*}$} & \textbf{0.74} & \textbf{2.5} \\
\hline\hline
\multicolumn{7}{p{1.05\columnwidth}}{\scriptsize $T_n$: noise temperature. $T_{\mathrm{phys}}$: physical operating temperature. $^{*}$average in-band value; $^{\dagger}$minimum in-band value; $^{\ddagger}$peak (maximum) value.} \\
\end{tabular}%
}
\end{table}

\section{Conclusion}
 
A 0.74~mW Ku-band cryogenic SiGe LNA in GF 130CBIC has been demonstrated. The $Q$-enhanced second-stage tank with a damping resistor and tunable $-g_m$ cell provides cryogenic stability and gain headroom. At 2.5~K, the LNA achieves an average noise temperature of 14\,K and a peak gain of 28~dB across 12--18~GHz, representing the lowest DC power reported for a Ku-band cryogenic SiGe LNA.

\section*{Acknowledgment}

The authors would like to thank the GF University Program for chip fabrication and Prof. Marvin Onabajo for the noise measurement equipment. This research was funded in part by the IEEE MTT-S Graduate Fellowship. The authors are also grateful to Jesse Balgley and Xuanjing Chu from the Columbia team for the van der Waals Josephson junction stacking.

\bibliographystyle{IEEEtran}
\bibliography{myrefs}

\begin{thebibliography}{10}
\providecommand{\url}[1]{#1}
\csname url@samestyle\endcsname
\providecommand{\newblock}{\relax}
\providecommand{\bibinfo}[2]{#2}
\providecommand{\BIBentrySTDinterwordspacing}{\spaceskip=0pt\relax}
\providecommand{\BIBentryALTinterwordstretchfactor}{4}
\providecommand{\BIBentryALTinterwordspacing}{\spaceskip=\fontdimen2\font plus
\BIBentryALTinterwordstretchfactor\fontdimen3\font minus \fontdimen4\font\relax}
\providecommand{\BIBforeignlanguage}[2]{{%
\expandafter\ifx\csname l@#1\endcsname\relax
\typeout{** WARNING: IEEEtran.bst: No hyphenation pattern has been}%
\typeout{** loaded for the language `#1'. Using the pattern for}%
\typeout{** the default language instead.}%
\else
\language=\csname l@#1\endcsname
\fi
#2}}
\providecommand{\BIBdecl}{\relax}
\BIBdecl

\bibitem{krinner_engineering_2019}
S.~Krinner, {\textit{et al.}}, and A.~Wallraff, ``{Engineering Cryogenic Setups for 100-Qubit Scale Superconducting Circuit Systems},'' \emph{EPJ Quantum Technology}, vol.~6, no.~1, p.~2, 2019.

\bibitem{CMOS_Quantum_System}
B.~Patra, {\textit{et al.}}, and E.~Charbon, ``{Cryo-CMOS Circuits and Systems for Quantum Computing Applications},'' \emph{IEEE Journal of Solid-State Circuits}, vol.~53, no.~1, pp. 309--321, 2018.

\bibitem{arute_quantum_2019}
F.~Arute, {\textit{et al.}}, and J.~M. Martinis, ``{Quantum Supremacy Using a Programmable Superconducting Processor},'' \emph{Nature}, vol. 574, no. 7779, pp. 505--510, 2019.

\bibitem{Review_quangtum}
N.~Ebrahimi and {\textit{et al.}}, ``Direct digital-to-physical synthesis: From millimeter-wave transmitter to qubit control,'' \emph{IEEE Microwave Magazine}, vol.~27, no.~5, pp. 50--69, 2026.

\bibitem{JPA}
T.~Yamamoto, {\textit{et al.}}, and J.~S. Tsai, ``{Flux-Driven Josephson Parametric Amplifier},'' \emph{Applied Physics Letters}, vol.~93, no.~4, p. 042510, 2008.

\bibitem{JPA_classical}
C.~Macklin, {\textit{et al.}}, and I.~Siddiqi, ``{A Near-Quantum-Limited Josephson Traveling-Wave Parametric Amplifier},'' \emph{Science}, vol. 350, no. 6258, pp. 307--310, 2015.

\bibitem{NbSe2}
J.~Balgley, {\textit{et al.}}, and K.~C. Fong, ``{Crystalline Superconductor-Semiconductor Josephson Junctions for Compact Superconducting Qubits},'' \emph{Physical Review Applied}, vol.~24, no.~3, p. 034016, 2025.

\bibitem{ref_zou2024}
Z.~Zou, {\textit{et al.}}, and J.~C. Bardin, ``{A 1.6-mW Cryogenic SiGe LNA IC for Quantum Readout Applications Achieving 2.6-K Average Noise Temperature From 3 to 6 GHz},'' \emph{IEEE Microwave and Wireless Technology Letters}, vol.~34, no.~6, pp. 753--756, 2024.

\bibitem{ref_SiGe_LNA_review}
J.~C. Bardin, ``{Silicon-Germanium Heterojunction Bipolar Transistors for Extremely Low-Noise Applications},'' Ph.D. dissertation, California Institute of Technology, 2009.

\bibitem{cryo_bjt}
X.~Jin, {\textit{et al.}}, and M.~Schröter, ``{Advanced SiGe:C HBTs at Cryogenic Temperatures and Their Compact Modeling With Temperature Scaling},'' \emph{IEEE Journal on Exploratory Solid-State Computational Devices and Circuits}, vol.~7, no.~2, pp. 175--183, 2021.

\bibitem{ref_wu2023}
R.~Wu and C.~Wang, ``{A 2--12 GHz SiGe BiCMOS Cryogenic LNA with 6 K Noise Temperature and 1.3 mW DC Power},'' in \emph{2023 IEEE MTT-S International Microwave Workshop Series on Advanced Materials and Processes for RF and THz Applications (IMWS-AMP)}, 2023, pp. 1--3.

\bibitem{Wong_2016}
W.-T. Wong, {\textit{et al.}}, and J.~C. Bardin, ``{A SiGe Ka-Band Cryogenic Low-Noise Amplifier},'' in \emph{2016 IEEE MTT-S International Microwave Symposium (IMS)}, 2016, pp. 1--3.

\bibitem{ref_montazeri2017}
S.~Montazeri and J.~C. Bardin, ``{A Sub-Milliwatt 4--8 GHz SiGe Cryogenic Low Noise Amplifier},'' in \emph{2017 IEEE MTT-S International Microwave Symposium (IMS)}, 2017, pp. 1--4.

\bibitem{ref_varonen2019}
W.~Ramírez, {\textit{et al.}}, and M.~Varonen, ``{Cryogenic Operation of a Millimeter-Wave SiGe BiCMOS Low-Noise Amplifier},'' \emph{IEEE Microwave and Wireless Components Letters}, vol.~29, no.~6, pp. 403--405, 2019.

\end{thebibliography}

\end{document}